# Optimizing MACD Trading Strategies: A Dance of Finance, Wavelets, and Genetics

Wangyu Chen[1], Zhenpeng Zhu[2]

1    Wangyu Chen,Nanjing Audit University,1006025254@qq.com

2    Zhenpeng Zhu, Huazhong University of Science and Technology,1958230983@qq.com

*    Correspondence: Wangyu Chen.1006025254@qq.com; Tel.: +86 18962931109

## Abstract

In today's financial markets, quantitative trading has become an essential trading method, with the MACD indicator widely employed in quantitative trading strategies. This paper begins by screening and cleaning the dataset, establishing a model that adheres to the basic buy and sell rules of the MACD, and calculating key metrics such as the win rate, return, Sharpe ratio, and maximum drawdown for each stock. However, the MACD often generates erroneous signals in highly volatile markets. To address this, wavelet transform is applied to reduce noise, smoothing the DIF image, and a model is developed based on this to optimize the identification of buy and sell points. The results show that the annualized return has increased by 5%, verifying the feasibility of the method.

Subsequently, the divergence principle is used to further optimize the trading strategy, enhancing the model's performance. Additionally, a genetic algorithm is employed to optimize the MACD parameters, tailoring the strategy to the characteristics of different stocks. To improve computational efficiency, the MindSpore framework is used for resource management and parallel computing. The optimized strategy demonstrates improved win rates, returns, Sharpe ratios, and a reduction in maximum drawdown in backtesting.

# Table of Contents

# 1. Introduction to the Problem

## 1.1. Problem Background

MACD (Moving Average Convergence Divergence) is a kind of important Average technical analysis indicators, for stocks, futures and other asset price action analysis. The formula for the calculation is

$$MACD = 2 \cdot (DIF - DEA)$$

where
DIF = short-term moving average (12 days) - long-term moving average (26 days) DEA = moving average of DIF (9-day)

When the MACD bar line is positive, the red line is long, and when the negative green line becomes long, it indicates a decline. An upward-crossed MACD of DIF is a buy, while a downward-crossed MACD of DIF is a sell. Can also be used to predict and deviate from the bear will deviate from the MACD, when the share price high than the previous one is lower than the previous highs higher and the MACD highs, cattle deviation appears, indicate stock will soon reverse declines; When the stock price is lower than the previous lows but MACD indicators low is higher than the previous low, a bear, predict stock price will rise soon.

However, since MACD is a medium and long term indicator suitable for long-term trend market, the shock market should be avoided and the reaction is lagging in the sudden rise, and the indicator cannot be effective. This paper aims to improve the MACD indicator to make it have better investment performance.

## 1.2. Problem Description

Based on your understanding of the context, the problem to be solved can be listed as follows:
·Cleaning and select useful data
·Set up the trading system and make it conform to the topic given rules when buying and selling MACD indicators, to each stock is given the odds, the odds, and frequency of the MACD indicators, and the corresponding income, annual yield, sharpe ratio, maximum retracement quantitative indicators, etc
·Establish a model to weaken the influence of wrong signals in the oscillating range
·Model and optimize the system so that it conforms to the bull deviation and bear deviation rules
·MACD parameters have different optimal parameters for different stocks, and methods should be sought to solve the most appropriate parameter ratio for their own stocks at this time
·improve the computational efficiency
To this end, we established a working outline, as shown in Figure 1:

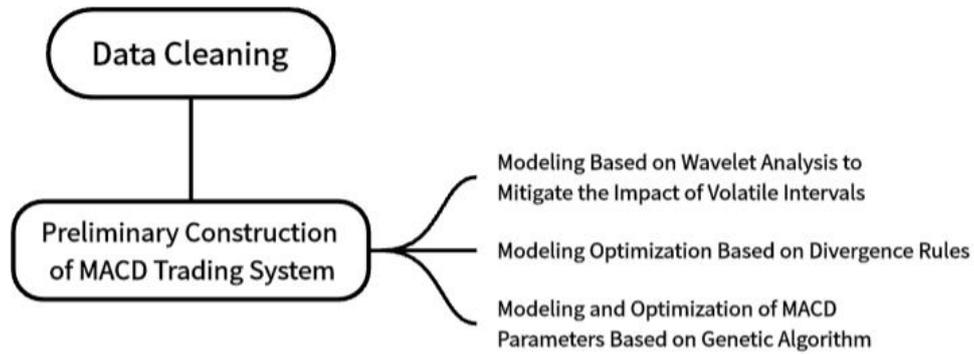

Figure 1 Outline of the work

## 2. Explanation of Symbols

| Symbols | Definition |
|---|---|
| $p_i$ | stock settlement price on day i |
| $P, P_{win}, P_{lose}$ | total profit and loss, gross profit, gross loss |
| $N, N_{out}, N_{win}$ | total number of trades in the investment, total number of sell trades in the investment, total number of profitable sell trades in the investment |
| $T$ | total time period |
| $V_f, V_i$ | sell value on day f of the investment, initial investment value on day i |
| $n$ | duration of the investment in years |
| $R_p$ | average return on the investment portfolio |
| $R_f$ | risk-free rate (average of the one-year government bond yield in China from May 2, 2012, to April 30, 2014, 2.653%) |
| $\sigma_p$ | standard deviation of the investment portfolio return |
| $a(i)$ | average value |
| $c(i)$ | situation where the stock price is within the range on a single day |
| $cs(i)$ | situation where the stock price is within the range for two consecutive days |
| $b_i$ | i-th level approximation (low-frequency part) in the wavelet |
| $d_i$ | i-th level detail (high-frequency part) in the wavelet |
| $h_k$ | low-pass filter coefficients of the wavelet |
| $g_k$ | high-pass filter coefficients of the wavelet |
| $\emptyset_1[f(t)]$ | DIF function after wavelet analysis |
| $\emptyset_2[f(t)]$ | traditional MACD value function |
| $P(x,y,z)$ | total profit and loss of the trading model in Problem 1, using the MACD indicator with parameters x, y, and z, where x, y, |

|  | and z are the short-term moving average value, long-term moving average value, and DIF moving average value respectively |
| --- | --- |
| elite_size | number of elite individuals in the genetic algorithm |
| elite | list of elite individuals in the genetic algorithm |
| fitness | list of individual fitness in the genetic algorithm |
| prob | list of selection probabilities for individuals in the genetic algorithm |
| pc | probability of crossover operation |
| point | randomly selected crossover point |
| $Gen_i[j]$ | j-th gene of individual i |

Table 1: Explanation of Symbols

# 3. Building and solving the model

## 3.1. Model Assumptions

Here are our model assumptions:
1. It is assumed that the basic situation of a stock can be measured from the dynamic change process of the closing price. When investors consider each investment choice, they rely on the probability distribution of security returns in a certain holding time.
2. Assume poundage influence can be neglected in the profit and loss.
3. Assume the principal is 500,000.
4. Assume both are a one-time buy and sell.

## 3.2. Build and improve your trading system based on MACD

### 3.2.1. Problem Analysis

We need on the basis of the MACD indicators to build a trading system, for each species is given the odds, odds and frequency of the MACD indicators, and the corresponding maximum gain, the annual interest rate, sharpe ratio, maximum retracement of quantitative indicators.

First we clean the received data set and then we compute it using the formula related to the indicator.

The limitations of MACD indicators have for turbulent session will give you the wrong trading signals and for certain hysteresis. Reading the literature, it can be seen that wavelet analysis can more realistically and synchronously depict the change law of the stock market [1], and better results can be obtained by replacing the MACD curve with the low-frequency part of the signal [2]. Based on the corresponding requirements, we use wavelet analysis to model the oscillating market and reduce the noise of DIF, so as to optimize the MACD calculation formula.

The background mentioned that the positive red line of the bar line becomes longer,

indicating that it is a rising trend, and the negative green line becomes longer, indicating that it is a falling trend. According to Jiang Yingshi's "A Brief Analysis of the application Rules and Limitations of the MACD Index" [3], we establish a model for the bar chart in the MACD index and put it into the code to judge the buying and selling situation.

In addition, there are cattle in the MACD deviation and bear, we according to the corresponding principle after improve the trading system to join the standard modeling formula makes the trading system more complete.

### 3.2.2. Data Preprocessing

First of all, we have collected from the Wind 510300. SH stock nearly 10 years of trade data such as, but not identical, we found that the different stock data information and the existence of containing too much 0, we has carried on the screen to such information, pick out the efficient code, date, closing price as the experimental information for analysis.

We obtained the results of all stocks in the code according to the following formula:

$$\text{win\_rate} = \frac{N_{win}}{N_{out}} * 100\%$$

$$\text{odds\_ratio} = \frac{\frac{P_{win}}{N_{win}}}{\frac{P_{lose}}{N_{out} - N_{win}}} * 100\%$$

$$\text{trade\_frequency} = \frac{N}{T} * 100\%$$

$$\text{total\_return} = \max(V_f - V_i)$$

$$\text{annual\_return} = (\frac{V_f}{V_i})^{\frac{1}{n}} - 1$$

$$\text{sharp\_ratio} = \frac{R_p - R_f}{\sigma_p} * 100\%$$

$$\text{max\_drawdown} = \max_{i,j:i<j}(\frac{p_j - p_i}{p_i} * 100\%)$$

### 3.2.3. Based on wavelet analysis modeling to mitigate the impact of the oscillation interval
#### 3.2.3.1. Mathematical modeling based on wavelet analysis

First, we model the determination of the oscillation range:

$$\begin{cases} a(p_i) = \frac{1}{10} \sum_{i}^{i+9} p_i \\ c(k) = \begin{cases} 1, & p_k \epsilon (a*0.985, a*1.015) \\ 0, & others \end{cases}, k\epsilon(i, i+9) \\ cs(k) = c(k) * c(k+1), k\epsilon(i, i+9) \end{cases}$$

If cs(k) = 1, then it is determined that the interval from k to k' no longer satisfies this condition and is considered an oscillation range. We have written a program to run this, as shown in Figure 2. The yellow-highlighted area indicates the oscillation range, where the MACD frequently gives erroneous signals, so this interval needs to be optimized.

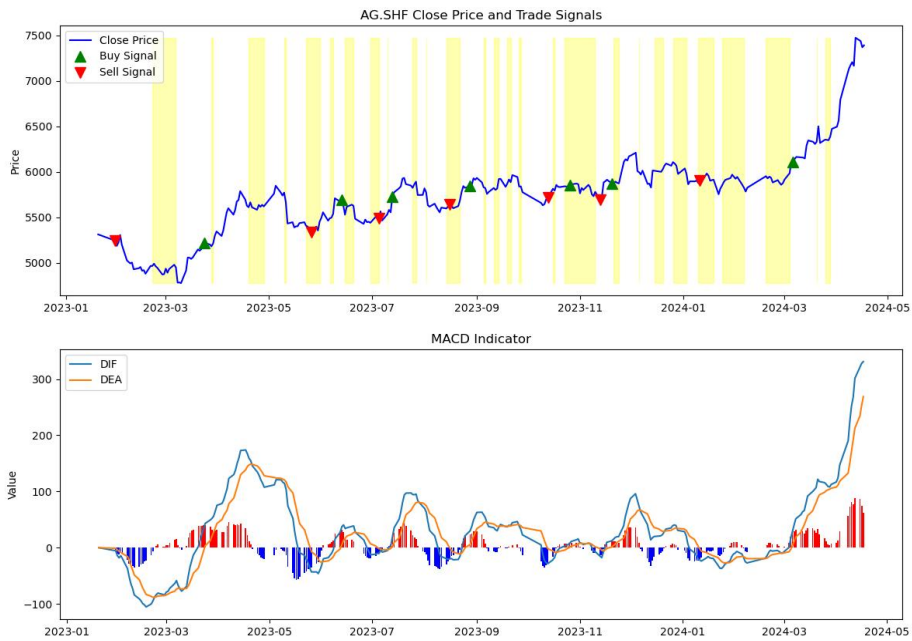

Figure 2: Oscillation Interval Determination

We draw on the discrete wavelet transform (DWT) formula from wavelet theory and use the Colf5 wavelet to perform a four-level wavelet decomposition of DIF. By repeatedly applying the wavelet and smoothing filters, the signal is progressively decomposed into low-frequency (approximation) and high-frequency (detail) components, resulting in a four-level low-frequency component $b_4$.

The original signal is denoted as f(t), and each level of wavelet decomposition can be represented by the following formula:
1. First-level Decomposition:

$$b_1(t) = \sum_k h_k f(t-2k), d_1(t) = \sum_k g_k f(t-2k)$$

2. Recursive Decomposition to the Next Level:
For j=2,3,4,…,

$$b_j(t) = \sum_k h_k\, b_{j-1}(t - 2k),\quad d_1(t) = \sum_k g_k\, b_{j-1}(t - 2k)$$

In this case, we recursively apply the above recurrence relation, using the approximation from the previous level as the input for the current level decomposition, resulting in:

$$b_4(t) = \sum_k h_k\, b_3(t - 2k)$$

Ultimately, the four-level low-frequency components extracted from the DIF signal using the Colf5 wavelet can effectively represent the fundamental (long-term) trend of the DIF signal.

Ultimately, the four-level low-frequency components extracted from the DIF signal using the Colf5 wavelet can effectively represent the fundamental (long-term) trend of the DIF signal.

The objective of the model is:

$$\max z = p_{t_2} - p_{t_1}$$

where the condition is:

$$\begin{cases} \emptyset_1[f(t)] = b_4(t) \\ \emptyset_2[f(t)] = \text{MACD} \\ t_2 \geq t_1 \geq 0 => t_1, t_2 \end{cases}$$

$t_1$ and $t_2$ are the buy and sell time points, respectively, that maximize $f(t_1)$-$f(t_2)$. After applying wavelet analysis and modeling, we have reduced the interference from high-frequency noise, thereby providing effective trading signals.

### 3. 2. 3. 2.  Model Empirical Testing

We conducted a wavelet analysis on the NR.INE stock and compared the results before (Figure 2) and after (Figure 3) the analysis. It can be observed that the buy and sell signals given by the MACD index became more accurate after the denoising process. The investment trading logic was modified to smooth the oscillation zone and successfully identify the mid-to-long-term trends. The buying and selling operations still followed the original criteria at the crossover points.

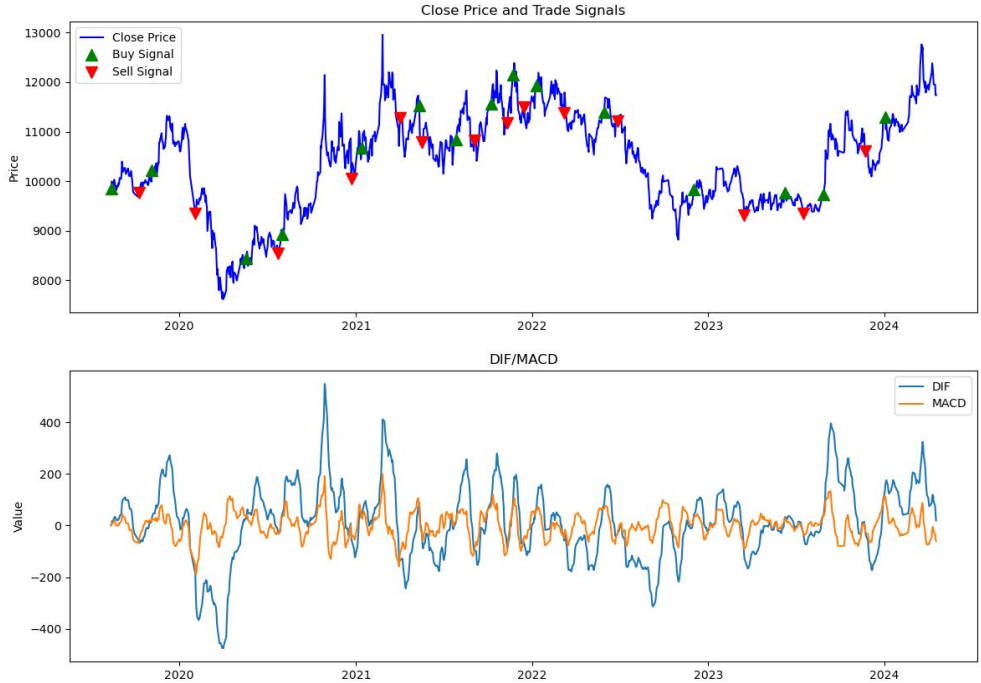

Figure 3: Buy and Sell Signals Before and After DIF Wavelet Analysis

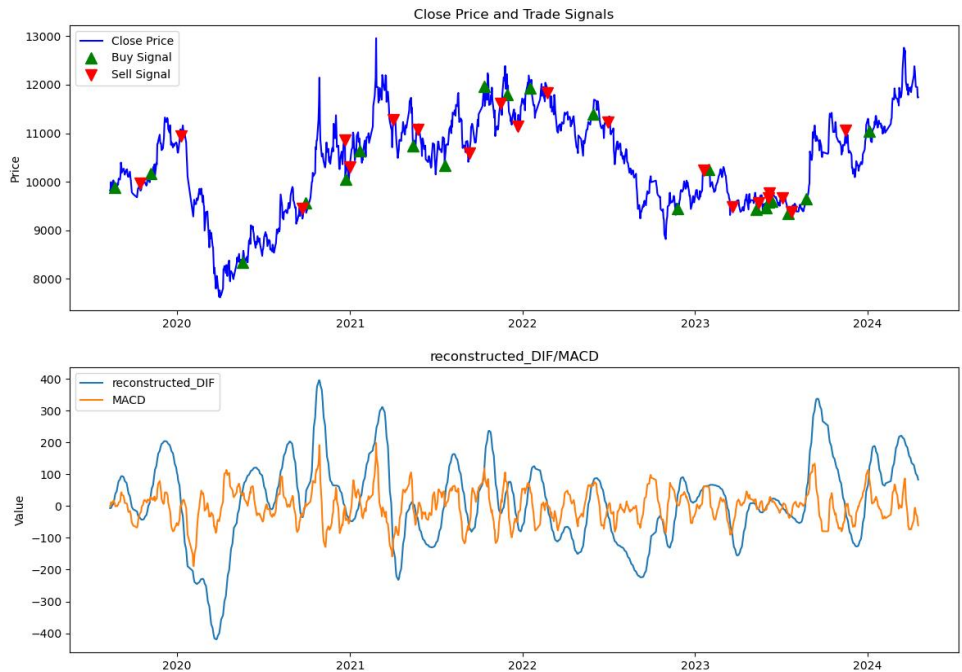

Figure 4: Buy and Sell Signals After DIF Wavelet Analysis

We selected several stocks and compared the index information before and after the modifications for empirical results analysis. It is evident that metrics such as winning rate, maximum returns, annualized return, and Sharpe ratio all increased significantly. Under the same investment, the returns soared, and the odds and maximum drawdown also showed some improvement. The quantification metrics of the top ten stocks before wavelet analysis are shown in Table 2 and Table 3:

| name | win_rate | odds_ratio | trade_frequency | total_return | annual_return | sharpe_ratio | max_drawdown |
|---|---|---|---|---|---|---|---|
| 510300.SH | 38.71% | 314.87% | 2.14% | 523467.64 | 6.44% | 105.82% | 20.22% |
| AG.SHF | 27.27% | 189.50% | 3.03% | -63352 | -1.17% | -49.79% | 48.10% |
| AL.SHF | 37.80% | 310.51% | 2.33% | 820655 | 3.53% | 69.83% | 24.70% |
| AP.CZC | 31.58% | 175.90% | 2.48% | -97253 | -3.50% | -27.19% | 51.41% |
| AU.SHF | 42.00% | 199.50% | 2.53% | 348633.18 | 3.43% | 48.66% | 27.95% |
| CF.CZC | 33.93% | 418.62% | 2.32% | 968325 | 5.77% | 124.89% | 28.45% |
| CU.SHF | 35.53% | 428.91% | 2.16% | 5489810 | 9.28% | 256.28% | 23.24% |
| EC.INE | 33.33% | 4291.77% | 3.77% | 714126.4 | 307.99% | 178.85% | 31.81% |
| FU.SHF | 39.66% | 186.25% | 2.44% | 695001 | 4.72% | 90.11% | 65.05% |

Table 2: Quantitative Metrics Before Wavelet Analysis

| name | win_rate | odds_ratio | trade_frequency | total_return | annual_return | sharpe_ratio | max_drawdown |
|---|---|---|---|---|---|---|---|
| 510300.SH | 70.00% | 211.74% | 2.76% | 1837111.74 | 14.37% | 267.15% | 14.04% |
| AG.SHF | 74.00% | 269.27% | 3.44% | 3184160 | 18.93% | 330.75% | 9.76% |
| AL.SHF | 71.96% | 212.87% | 3.04% | 7764640 | 10.55% | 374.89% | 18.40% |
| AP.CZC | 45.83% | 267.35% | 3.13% | 812004 | 17.20% | 173.48% | 24.82% |
| AU.SHF | 67.74% | 355.18% | 3.14% | 2815893.58 | 12.81% | 341.85% | 8.03% |
| CF.CZC | 64.79% | 668.54% | 2.94% | 5088665 | 13.40% | 342.54% | 12.91% |
| CU.SHF | 66.29% | 388.24% | 2.53% | 43595900 | 17.37% | 501.26% | 8.99% |
| EC.INE | 100.00% | 0.00% | 3.77% | 1219231.6 | 608.09% | 236.62% | 30.45% |
| FU.SHF | 77.33% | 387.32% | 3.15% | 65875156 | 29.52% | 439.48% | 13.99% |

Table 3: Quantitative Metrics After Wavelet Analysis

### 3.2.4. Trading Optimization Based on Divergence Principle

When the price high is higher than the previous high but the MACD high is lower than the previous high, it is called a Bullish Divergence. This indicates that the stock is likely to reverse and decline soon. When the price low is lower than the previous low but the MACD low is higher than the previous low, it is called a Bearish Divergence, which suggests that the stock is likely to reverse and rise soon. To improve the MACD-based trading system, we introduced a new modeling condition based on these divergence principles:

$$\text{Bullish Divergence} \begin{cases} p_{high} > max(p_{prev}, p_{next}) \\ p_{max} > max(p_{15days}) \\ p_{high} > p_{prevhigh} \\ MACD_{high} < MACD_{prevhigh} \end{cases}$$

$$\text{Bearish Divergence} \begin{cases} p_{low} < min(p_{prev}, p_{next}) \\ p_{low} < min(p_{15days}) \\ p_{low} < p_{prevlow} \\ MACD_{low} > MACD_{prevlow} \end{cases}$$

We extracted stocks and used Python code to identify Bullish Divergence (as shown in Figure 5) and Bearish Divergence (as shown in Figure 6):

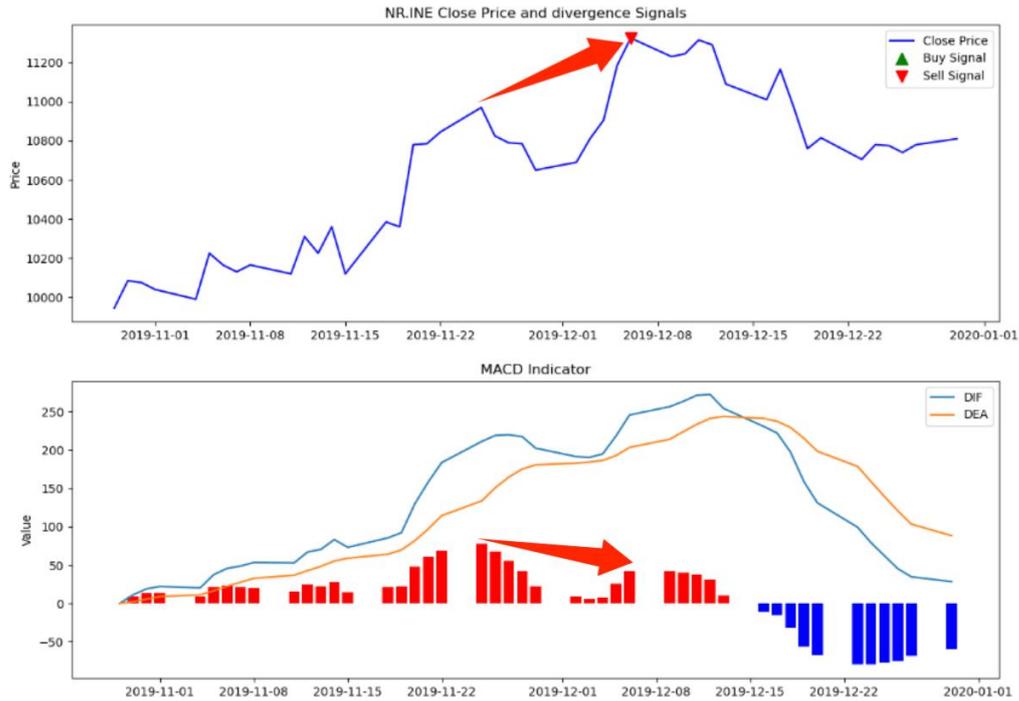

Figure 5: Bullish Divergence Description

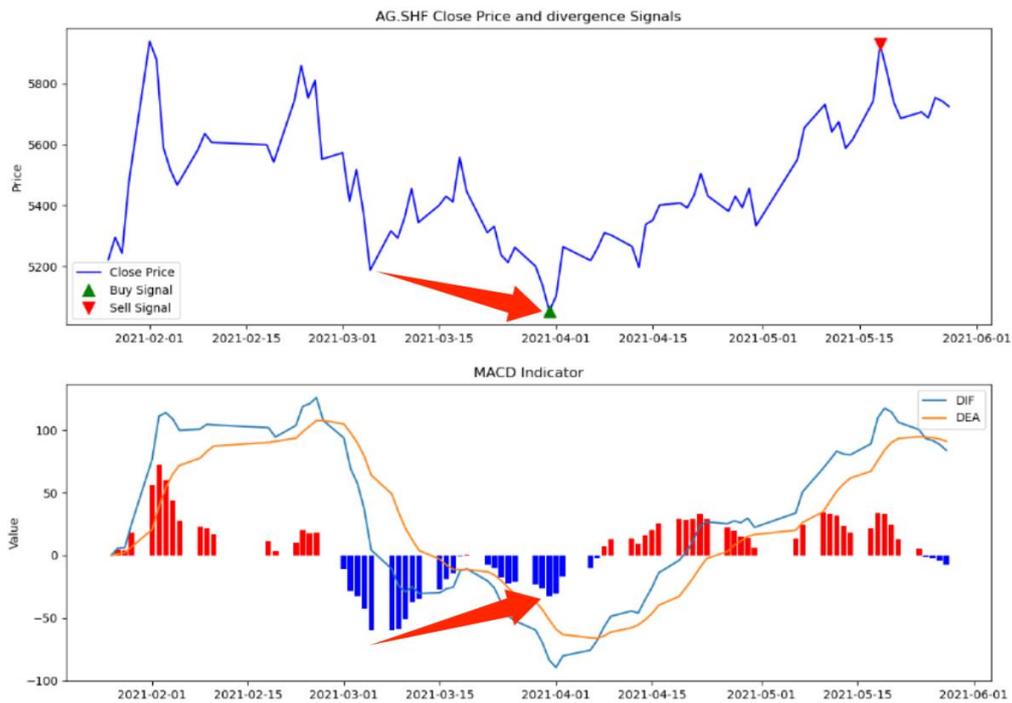

Figure 6: Bearish Divergence Description

### 3.2.5. Model Backtest Results

After establishing and optimizing the system, we compared the original data (Table 4) with the result data (Table 5). We found that after optimizing the MACD indicator, we achieved relatively higher returns, thus completing the construction of the trading system.

| name | win_rate | odds_ratio | trade_frequency | total_return | annual_return | sharpe_ratio | max_drawdown |
|---|---|---|---|---|---|---|---|
| 510300.SH | 38.71% | 314.87% | 2.14% | 523467.64 | 6.44% | 105.82% | 20.22% |
| AG.SHF | 27.27% | 189.50% | 3.03% | -63352 | -1.17% | -49.79% | 48.10% |
| AL.SHF | 37.80% | 310.51% | 2.33% | 820655 | 3.53% | 69.83% | 24.70% |
| AP.CZC | 31.58% | 175.90% | 2.48% | -97253 | -3.50% | -27.19% | 51.41% |
| AU.SHF | 42.00% | 199.50% | 2.53% | 348633.18 | 3.43% | 48.66% | 27.95% |
| CF.CZC | 33.93% | 418.62% | 2.32% | 968325 | 5.77% | 124.89% | 28.45% |
| CU.SHF | 35.53% | 428.91% | 2.16% | 5489810 | 9.28% | 256.28% | 23.24% |
| EC.INE | 33.33% | 4291.77% | 3.77% | 714126.4 | 307.99% | 178.85% | 31.81% |
| FU.SHF | 39.66% | 186.25% | 2.44% | 695001 | 4.72% | 90.11% | 65.05% |

Table 4: Quantitative Indicators Before Adding Divergence Judgment

| name | win_rate | odds_ratio | trade_frequency | total_return | annual_return | sharpe_ratio | max_drawdown |
|---|---|---|---|---|---|---|---|
| 510300.SH | 74.55% | 91.58% | 5.53% | 1146414.992 | 10.93% | 233.02% | 30.01% |
| AG.SHF | 80.65% | 109.29% | 6.03% | 1902789 | 14.60% | 252.11% | 13.44% |
| AL.SHF | 73.33% | 62.01% | 5.15% | 842270 | 3.59% | 73.72% | 31.35% |
| AP.CZC | 68.75% | 129.07% | 6.46% | 544392 | 12.88% | 129.05% | 17.55% |
| AU.SHF | 84.15% | 133.89% | 5.31% | 2589568.12 | 12.30% | 380.24% | 9.50% |
| CF.CZC | 76.40% | 81.52% | 5.07% | 1418935 | 7.26% | 182.60% | 19.53% |
| CU.SHF | 76.92% | 68.64% | 4.87% | 2800590 | 6.98% | 197.38% | 32.49% |
| EC.INE | 100.00% | 0.00% | 3.77% | 1219231.6 | 608.09% | 236.62% | 30.45% |
| FU.SHF | 76.09% | 55.37% | 5.06% | 4125565 | 12.49% | 214.16% | 34.75% |

Table 5: Quantitative Indicators After Adding Divergence Judgment

### 3.3. Improving the MACD Indicator to Optimize the Investment Performance of the Trading System

#### 3.3.1. Problem Analysis

Based on the insights we have gathered, we have made initial efforts to establish and optimize the trading system. However, after reviewing the relevant materials, we discovered areas that still require improvement. We found that the commonly used MACD parameters are not necessarily the most suitable for the specific stock. In Li Yitong's study, "Parameter Optimization Research for the Quantitative Trading Model of Shanghai Gold Futures Based on MACD Indicator" [4], exhaustive algorithms and genetic algorithms were used to improve the traditional MACD parameters, leading to higher returns. We adopted this method to model the genetic algorithm, and through continuous adjustments in the dataset, we aimed to find the most suitable parameter values for this specific stock.

#### 3.3.2. Optimizing MACD Parameters Through Genetic Algorithm Modeling

Genetic algorithm is an optimization algorithm based on the theory of biological evolution, used to solve search and optimization problems. It is a heuristic algorithm that gradually improves the quality of solutions by simulating the genetic evolutionary mechanisms in biology, including selection, crossover, and mutation, to find the optimal solution or an approximate optimal solution.

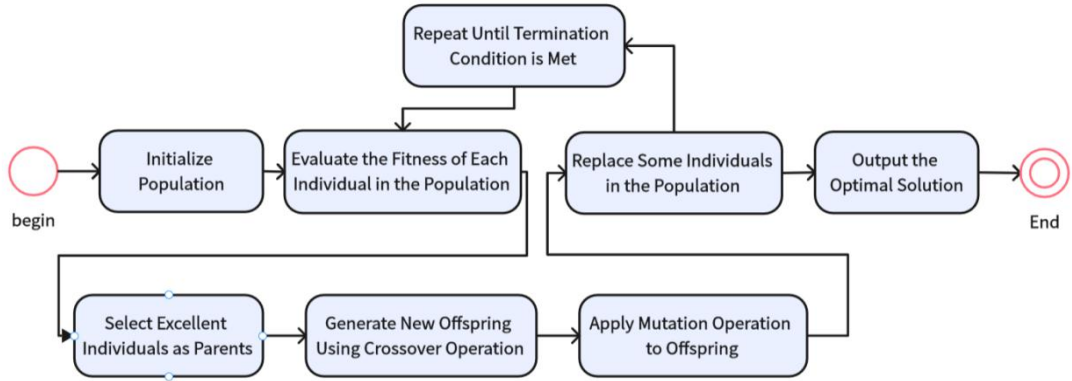

Figure 7: Genetic Algorithm Flowchart

We proceed with the following modeling based on the algorithm flowchart:

Let x, y and z represent the short-term moving average, long-term moving average, and the moving average of DIF for the MACD indicator, respectively. Then:

1. P(x,y,z) represents the total profit and loss value of the MACD indicator with parameters x, y, and z, which is applicable to the trading model mentioned earlier.

2. a,b and c are the optimal MACD parameters for a single stock.

Model objective：

$$a, b, c = argmax_{x,y,z}(P(x, y, z))$$

Constraints:

$$\begin{cases} Fast\ line\ period\ x \in [5,20] \\ Slow\ line\ period\ y \in [20,50] \\ Signal\ line\ period\ z \in [5,25] \\ \quad \Delta Parameters \in Z \end{cases}$$

### 3.3.2.1. Evaluating the Fitness of Each Individual in the Population

We substitute the MACD parameters x,y,z for each individual into the model built in the previous section (Section 1) and calculate the total profit and loss value P(x,y,z).This value is then used to evaluate the fitness of the individuals in the population.

### 3.3.2.2. Selecting Elite Individuals as Parents

First, we adopt an elite strategy, a commonly used method in optimization algorithms, where the individuals with the highest fitness are selected to preserve high-quality genes and accelerate the convergence of the algorithm. We control the elite size to be 1/10 of the total population, i.e., we retain 51 elite individuals to ensure that superior genes are passed on to the next generation. Below is the specific mathematical modeling process:

We define:
- elite_size as the number of elite individuals
- indices$_{elite}$ as the index list of selected elite individuals
- elite as the list of selected elite individuals

$$elite\_size = max(1, \frac{num\_individuals}{10})$$
$$indices_{elite} = argsort(fitness)[-elite\_size:]$$
$$elite = population[indices_{elite}]$$

Next, we select individuals probabilistically from the remaining population based on fitness, in order to form the new population. We first build a model to compute the adjusted fitness index:

$$max\_fitness = max(fitness)$$
$$exp\_v = e^{fitness - max\_fitness}$$

Where max_fitness is the maximum value in the fitness list,and exp_v is the fitness list optimized using the exponential function. The subtraction of the maximum fitness max_fitness is done for numerical stability, preventing overflow caused by exponential growth.

To avoid potential negative values for P(x,y,z) that could prevent the calculation of probabilities, we apply the exponential function, ensuring all probabilities are positive. To guarantee the sum of the probabilities equals 1, we use the softmax function to model the selection probabilities. The probability of an individual being selected is proportional to the adjusted fitness index of that individual divided by the sum of the adjusted fitness indices of all individuals. The modeling result is as follows:

$$prob_i = \frac{exp\_v_i}{\sum exp\_v_i}$$

Where $prob_i$ is the probability of each individual being selected.

Finally, we select individuals based on their probabilities, choosing all individuals except the elites, and combine them with the elite individuals to form a new population:

$$chose\_indice = random\_choice(num\_individuals, siz = num\_individuals - elite\_size)$$
$$new\_population = concatenate(population[chosen\_indices], elite)$$

### 3. 3. 2. 3.  Using Crossover Operation to Generate New Offspring

The main purpose of the crossover operation is to combine gene segments from two parent individuals to create new offspring, thus increasing the genetic diversity of the population. This operation is inspired by the biological process of sexual reproduction, helping the algorithm explore new areas in the solution space and preventing it from getting trapped in local optima.

The most common crossover method in genetic algorithms is Single-Point Crossover. This method selects a random point (i.e., the crossover point) in the gene sequence of an individual, then exchanges the gene segments after the crossover point between the two parent individuals.

Let's define:
- pc (Crossover rate) : the probability of performing a crossover operation.

- point: the randomly selected crossover point, within the range [1, 2].
- $gen_i[j]$ : the j-th gene of individual i.

The crossover process model is as follows:

$$gen_i'[j] = \begin{cases} gen_i[j], & j < \text{point} \\ gen_{i+1}[j], & j \geq \text{point} \end{cases}$$

$$gen_{i+1}'[j] = \begin{cases} gen_{i+1}[j], & j < \text{point} \\ gen_i[j], & j \geq \text{point} \end{cases}$$

For every pair of individuals (i, i+1) starting from 0, if the random number is smaller than the crossover rate pc, the gene segments after the crossover point are exchanged, generating two new offspring.

### 3. 3. 2. 4. Using Mutation Operation to Mutate the Offspring

The purpose of the mutation operation is to introduce randomness by making small random changes to some genes of an individual, thereby increasing the genetic diversity of the population. This operation helps the algorithm escape from local optima and explore uncharted areas in the solution space.

Let's define the mutation rate:

pm (Mutation rate): the probability that each gene will undergo mutation.

Assume that each gene has a range bounds[i], where bounds[i][0] is the minimum value and bounds[i][1] is the maximum value.

The probability of mutation operation is pm. For each gene i:

$$gen_i = \begin{cases} \text{randint}(bounds[i][0], bounds[i][1]), & \text{rand}() < pm \\ gen_i, & \text{otherwise} \end{cases}$$

Here, rand() is a function that returns a random floating-point number in the range [0, 1), and randint(a, b) is a function that returns a random integer in the range [a, b].

### 3. 3. 2. 5. Combining the New Population

After performing the crossover and mutation operations, the new individuals generated through the compilation and price-difference operations replace the original individuals, forming a new population. This entire process is typically repeated in the main loop of the genetic algorithm until the maximum fitness value (max_fitness) remains unchanged for 8 consecutive generations, indicating full convergence.

### 3. 3. 2. 6. Optimization of Computation Using MindSpore Framework

In practical applications, we found that the genetic algorithm had a relatively slow computation speed. Therefore, we chose to introduce the optimization framework of MindSpore to accelerate the computational process of the genetic algorithm, particularly during the fitness evaluation and individual selection stages. MindSpore's automatic differentiation and parallel computing capabilities enabled the genetic algorithm to complete multiple iterations and large-scale searches in a shorter time, significantly improving the algorithm's efficiency.

Additionally, MindSpore's hardware acceleration function further reduced the runtime when handling complex data and computational tasks, ensuring that the algorithm could still perform efficiently under resource constraints. This, in turn, enhanced the accuracy of the experimental results and the effectiveness of the optimization.

### 3.3.3. Empirical Results of the Model and Investment Trading Logic Analysis

We attempted to use the genetic algorithm to address the mismatch between the MACD parameters of individual stocks and the conventional parameter values. The optimized parameters for the stock 510300.SH were found to be [9, 22, 25]. By comparing this result with the outcome from the first part of the analysis, we observed significant improvements in win rate, Sharpe ratio, and returns, which increased by 13%, 86%, and 272%, respectively. Additionally, the odds and maximum drawdown decreased by 36% and 12%, respectively. The specific values are shown in Table 6:

| MACD Parameter Values | [12,26,9] | [9,22,25] |
| --- | --- | --- |
| Win Rate | 74.55% | 84.21% |
| Odds | 91.58% | 58.81% |
| Frequency | 5.53% | 6.63% |
| Max Profit | 1146414.99 | 4267080.92 |
| Annualized Return | 10.94% | 21.69% |
| Sharpe Ratio | 233.02% | 433.76% |
| Max Drawdown | 30.01% | 26.50% |

Table 6: Genetic Algorithm Optimization Results(based on MindSpore framework)

We plotted and summarized the buy points, and found that the optimization influenced the trading logic. By adjusting the fast line, slow line, and signal line periods according to market conditions, the optimization reduced noise and false signals, improving the quality of trading signals and making the trading logic more aligned with the characteristics of the stock market itself. In addition, the wavelet analysis introduced earlier also made the trading logic more compatible with the long-term characteristics of the MACD indicator. The trading logic now tends to focus on analyzing long-term trends to determine the timing for buying and selling, purifying short-term noise signals which the MACD is less effective at handling. The divergence optimization enriched the MACD's trading logic by allowing it to predict large upward or downward trends, making additional trading decisions based on these predictions.

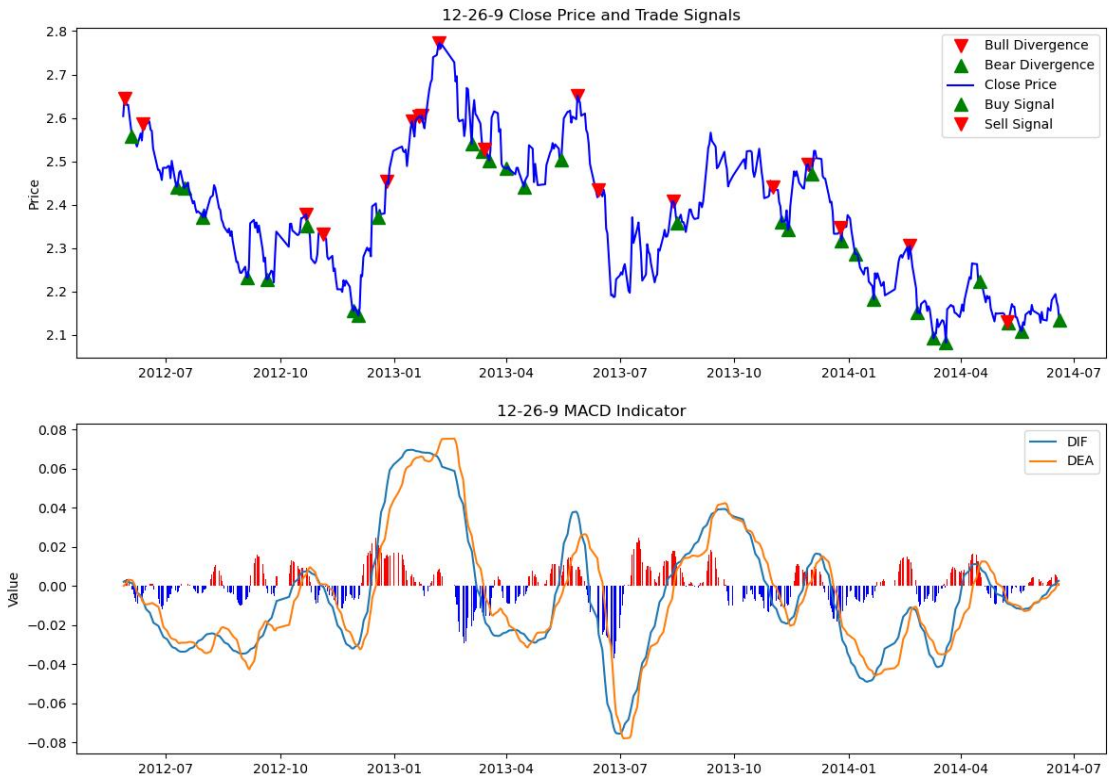

Figure 8: Trading Chart with Original MACD Indicator Parameters(based on MindSpore framework)

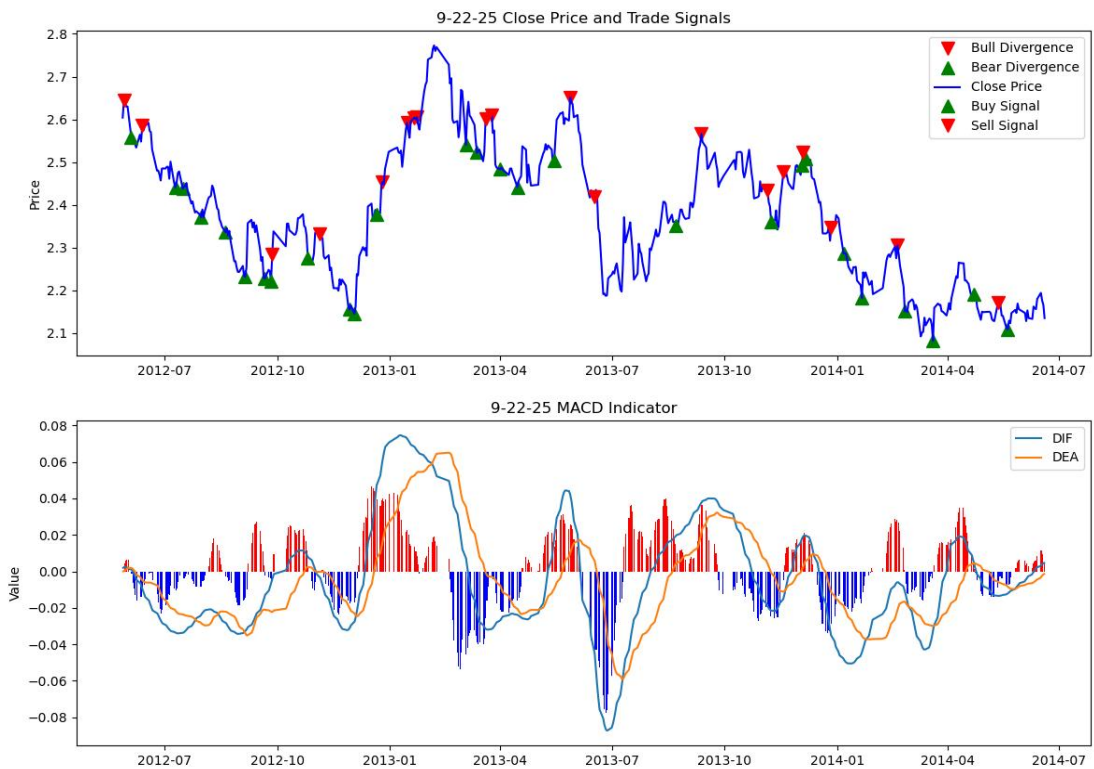

Figure 9: Trading Chart with Modified MACD Indicator Parameters(based on MindSpore framework)

# 4. Model Evaluation

| Advantages | Disadvantages |
|---|---|
| 1. The model in this paper addresses the inaccurate identification of oscillation zones, and the use of wavelet analysis effectively reduces the probability of frequent false signals in these zones.<br>2. The application of the divergence principle enhances the accuracy of the buy and sell signals generated by the MACD indicator.<br>3. The modeling of the MACD parameters using genetic algorithms enhances the adaptability of the trading system to individual stocks<br>4. The use of the MindSpore framework in this paper improves computational efficiency and accelerates data mining. | 1. The buy/sell signal strength needs to be weighted to improve the signal accuracy. The optimization of the buying strategy has not yet been completed in this paper.<br>2. The issue of lag in the MACD indicator's ability to identify sharp increases and decreases has not been solved. A compatible solution is yet to be found that both maintains the long-term effectiveness of the MACD indicator and addresses short-term identification issues.<br>3. The genetic algorithm can be further optimized by separating the buy and sell signals for more detailed adjustments. |

Table 7: Model Evaluation Results

# 5. Conclusion

In this paper, we made three improvements to the trading system based on the MACD indicator through variable control:

1.We used the principle of wavelet analysis to mathematically model stock market volatility charts. After multiple iterations of the model, we obtained a smoothed stock market chart that better aligns with the judgment logic of medium-to-long-term MACD indicators. This successfully reduced false signals, making the entire stock movement process smoother.

2.We improved the buy and sell signals generated by the MACD indicator under other conditions based on the principle of divergence, enriching its functionality.

3.We applied the genetic algorithm under the MindSpore framework for iterative selection, which quickly converges to the optimal solution. The genetic algorithm, which is adapted from biology to the stock market, selects the most suitable parameters for individual stock MACD indicators, forming a more practical MACD model with specific applicability.

Through the construction and optimization of the trading system, we have improved the strategy of the MACD-based trading system to a certain extent.

## 6. Acknowledgments

We sincerely thank the MindSpore team for providing us with powerful technical support and the MindSpore community for offering this experimental opportunity and platform. The efficient computing framework of MindSpore not only helps us accelerate the processing speed of computing tasks, enabling us to fully utilize hardware acceleration devices, but also provides flexible resource management functions, making the computing process in this research more efficient and stable.

Next, we would like to express our gratitude to our families. Whether it is the challenges encountered during the research process or the trivial matters in daily life, your support has been our source of motivation and also the solid foundation that helps us complete this work.

Finally, we would like to thank everyone who has provided us with assistance on our academic path. Your support and encouragement have enabled us to successfully complete this work.

The road ahead is still full of challenges, but we will continue to strive and live up to your care and expectations.

## 7. References

[1] 王哲, 王春峰, 顾培亮. 小波分析在股市数据分析中的应用[J]. 系统工程学报, (03):286-289+295, 1999.
[2] 袁修贵, 侯木舟. 小波分析在证券分析中的应用[J]. 中南工业大学学报(自然科学版), (01):103-106, 2002.
[3] 蒋莹诗. 浅析MACD指标的应用法则及其局限性[J]. 物流工程与管理, 36(06):149-150, 2014.
[4] 李易桐. 基于MACD指标的沪金期货量化交易模型的参数优化研究[D]. 广西大学, 2018.